# Impact Analysis of Reducing Risks Concerning the Informatics Systems

AUTHORS: Floarea BAICU  - Hyperion University
Maria Alexandra BACHES - CONSIS PROIECT

**Abstract**
In this paper are presented methods of impact analysis on informatics system security accidents, qualitative and quantitative methods, starting with risk and informational system security definitions.
It is presented the relationship between the risks of exploiting vulnerabilities of security system, security level of these informatics systems, probability of exploiting the weak points subject to financial losses of a company, respectively impact of a security accident on the company.
Herewith are presented some examples concerning losses caused by excesses within informational systems and depicted from the study carried out by CSI.

**1. Introduction. Problem statement**

The International Standard ISO/CEI 17799 [1] *Informatics technology – Security techniques – Code of practice for Informatics security management* defines **risk** as the combination between the probability of occurrence of certain event and its consequences. From the point of view of informatics systems' security, the risk represents the combination between the probability of occurrence of a break in the security system of informatics systems and the impact on the capacity of those systems for carrying out the designed security functions.

The risk can be considered like as a threat that could exploit the possible vulnerabilities of the system, with a certain probability. It is an undesired and unpleasant event, which waits to occur, but that due certain reasons may not appear or through certain methods may be avoided. In order to prevent the occurrence of an undesired event generating considerable impact on the security of informatics systems, security measures should be taken. These security measures are called simply **measures** or **controls**. In this article we are not talking about the security measures that can be adopted in order to reduce the risk level concerning the security.

**Risk level** is an arbitrary indicator, denoted L, which allows grouping certain risks into equivalency classes. These classes include risks which are placed between two limit levels –  acceptable and unacceptable – conventionally established. It is determined through risk evaluations, on the basis of an adequate combination between the occurrence probability of a security event and the maxim consequences (impact) that the event may have upon the respective system.

**The acceptable risk level** is the risk level conventionally admitted by the organization management, regarded as not causing undesirable impacts on its activity. This is determined by methodic and specific evaluations.

**The residual risk** is considered to be the reminder after the risk treatment. As a general rule, the residual risk may be regarded as risk on acceptable level.

Assuming the risk definition set forth upwards, this is a positive real number $R$ which may be represented by the area of a quadrangle surface, having as one side the **P**robability of occurrence of certain event, noted with $P$, and as the other side the consequences of that event occurrence, respectively the **I**mpact of the undesirable event, noted with $I$, upon the security of the studied organization. Mathematically speaking, the same area may be obtained through various combinations between $P$ and $I$, of which the preferred one is quite the product between probability and impact. There are a lot of **P**robability – **I**mpact couples generating the same **R**isk $R$, defining quadrangles of same area as illustrated in figure 1.

If the vertexes of such quadrangles, which are not on axes, are linked through a continuous line it results a **hyperbolic curve** $C$, named the **Curve of Risk** [2]. This curve allows the differentiation between the **acceptable risk** (Tolerable – T) and the **unacceptable** one (Non-Tolerable – NT).

Thus, the risk of occurrence of a certain event A, with high impact, with serious consequences but low probability of occurrence, defined by coordinates placed below the represented acceptability curve is considered **acceptable**, while the risk of event B, with less serious consequences but high probability of occurrence, of which coordinates are placed upwards the curve, is considered **unacceptable**.

Hyperbolic curve of risk based on couples (**P**robability, **I**mpact) is illustrated in figure 1.

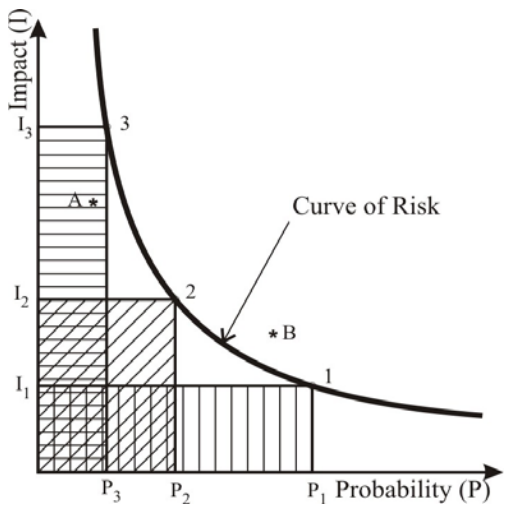

**Figure 1.** Graph representation for equivalency of Risks defined by different Probability – Impact couples

**Security function** is a function of a system (equipment) by help of which the risk is eliminated or reduced, or is only drawn the attention of a risk occurrence.

**Security level** is an indicator showing in general the security status of a system. It is determined indirectly, by determining the risk level and it is in inverse proportion to it.

The following relation describes the connection between the risk level and the security level:

$$Security = f(Risk) = \frac{1}{Risk} \quad (1)$$

and is presented in the table 1.

Concerning risk levels, in the present paper, due to practical reasons, for significations present in column 2 we will use the abbreviations present in column 3 of table 1.

**Table 1**: Relation between the risk level and the security level

| No. | Risk Level | Risk signification | Notation | Security Level | Security signification |
|---|---|---|---|---|---|
| 0 | 1 | 2 | 3 | 4 | 5 |
| 1 | $R_1$ | Minimum, Non risk | N | $S_7$ | Excellent |
| 2 | $R_2$ | Very low | VL | $S_6$ | Very good |
| 3 | $R_3$ | Low | L | $S_5$ | Good |
| 4 | $R_4$ | Medium | M | $S_4$ | Acceptable |
| 5 | $R_5$ | High | H | $S_3$ | Low |
| 6 | $R_6$ | Very high | VH | $S_2$ | Very low |
| 7 | $R_7$ | Critical | C | $S_1$ | Insignificant |

**Risk criteria** are reference terms against which the meaning of the risk (level) is determined.
There are several risk criteria categories, such as for example:
- specific consequences;
- costs or associate benefits;
- socio-economical aspects;
- the perception of interested parties;
- the occurrence frequency of security incidents;
- the cumulated effects of some incidents occurrence;
- uncertainty rate of determined risk level and of accepted trust level;
- the residual risk level admitted by each organization.

### 2. Economic improvement of the Security Level

All risk assessment methods present some drawbacks. Basically these are as follows:
- the values utilized are estimative;
- calculations are based on statistic and probability analyses;
- the data must be periodically up-dated.

A reduced security level has as effect the increase of business risk, as presented in table 1 Ensuring a high level of security can affect the business with significant expenses that can be utilized for other purposes, considered by many much more important.

Like in other activities here too must be found the optimum balance between the costs and benefits. It may be considered that the law 20-80 (Pareto Diagram) can be applied also in this case, with satisfactory results [2,3]. It is also considered that in most cases, 20% of the costs are reflected in accomplishments of desired benefits in a percentage of 80% - in our case minimization of the security costs. A maximum security $S_7$ (respectively an increase of only 20%) can be obtained with addition of extremely high expenses, practically with 80%. In figure 2 is shown the graphic illustration of this law concerning relation costs-benefits in order to achieve security of informatics systems.

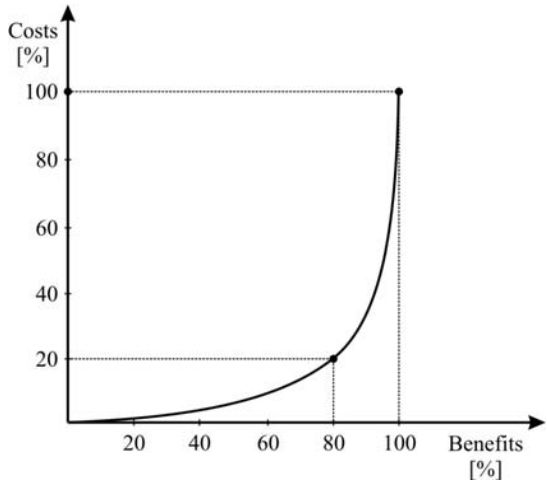

**Figure 2.** Relationship costs - benefits.

The manager of the organization is the person deciding how much is he willing to pay for the security, decision that is taken on the basis of a technical-financial analysis.

Security is very difficult to quantify. One can never state that within the organization we have a certain level of security. It only can be estimated as an excellent, very good, good, acceptable, low or insignificant level of security. Even though, an evaluation (at least financial) can be made concerning the security level..

Always implementation or tests and improvement the security generate costs concerning equipments, personnel and time spent for implementing security, including checks or simulations Those risk assessment are carried out mainly in the big companies and eventually in the medium-size ones. In small organization there is no enough personnel trained to do that kind of analysis and also there is no money to pay a specific company to do it. Taking all that into account, the least they could do is to spend some money for a minimum set of security measures. It is well known that managers do not invest in things that are not profitable at that moment and that, when they do spend different amounts of money, those are under the imposed limit for that kind of costs. In such case, there must be assured a security and its costs has to be under the line of a certain amount of money.

In order to measure the security level within a company with concern to means utilized for automatic data processing there can be utilized a *financial Security Indicator*, noted as $SI_f$, which is defined by the following formula:

$$SI_f = \frac{Ce + \sum_i^n P_i \cdot Cc_i}{Ce} - 1 \qquad (2)$$

where: Ce – the cost of calculation technique equipment and software used;

Pi – the weight of the measure agreed in the security system;

Cc – the cost of the controls for the equipment or software adopted.

In the case in which there are no investments in equipment assuring security the value of this indicator is zero.

When $0 < \text{SI}_f \leq 1$ – there is a minimum security level, but not inexistent.

When $\text{SI}_f > 1$ then we can say that the way in which it is assured the security is more expensive than the equipment itself. In this case, the reasons are several, such as:
- the risks were not well evaluated and/or security measures were exaggerated;
- the equipment (computer) is not of quality and needs additional equipments;
- the up-dated equipment value is low as compared to the cost of security measures.

In the first case, not to lose information, there has been made an exaggeration about the security measures – like financial investment or like number. In other words, the prices for those investments are too expensive or it is invested more than it is needed.

The second case occurs when there are bought poorly-qualitative computers and later equipment that may compensate this situation are required. This is the case of computers "no names" or even if they have, they are old and no longer present safety under operation. It is quite often encountered the situation when new bought computers get blocked in case of electric voltage fluctuations.

The third case is encountered when the cost of equipment is up-dated to the new value. If the computers as well as the control equipments were bought at close data and the cost of the equipment as well as that of control measures are simultaneously up-dated, then the $\text{SI}_f$ does not change so much.

When is desired an optimization formula for the software price, then we have to consider that the license to be bought can be utilized for each personal computer (PC) to be installed on and that only certain operating soft beneficiate of installation licenses only on the server serving these PC.

There will be taken into account the costs for all software installed on the respective computer, up-date cost for each of it, possibility of a breakdown for the hard disk and need to reinstall the program on that computer, cost of antivirus to be installed and its permanent updating. All these special soft have various prices. For example a Windows XP 2003 license, type OEM (can be installed only one time) costs about 150€ and a license that can be reinstalled, for the same software product, exceeds 350€. There must be carried out an analysis that should consider the probability of PC breakdown, due to several reasons, requiring reinstallation about 2 (two) times within 3 (three) years – period of time agreed as moral wear of this software, after which a re-installable license is purchased.

The price for an AutoCAD license exceeds 3500 € and it can be reinstalled on any computer, subject that the program is not running on several computers simultaneously. The acquisition of "subscription" option for the respective AutoCAD costs approximate 220 € per year and allows an automatic up-date to the versions to be created.

There is also the opportunity to "rent" AutoCAD license, for a certain period (for example 3 years), for 40-60 € per year for each PC, an offer that is starting to be successful.

All these economical analyses are part of the organization development strategy, representing an independent and difficult actions.

**3. Economically analysis of the impact reduction**

There are several methods for estimating the possible losses and the calculation of recovery costs.

An already classical method for economical analysis of the threat impact mentioned in ISO/IEC 13335-3 *Informatics technology- Guidelines for the management of IT security – Part 3: Techniques for the management of IT security* is called **A**nalyzed **L**oss **E**xpectancy – ALE. It is a simple quantitative method, which allows estimation of the possible losses based on already existent company records, kept by an expert, in the organization.

ALE depends on the value of the asset, on the possibility to loose its value, vulnerability to the respective threat and threat occurrence frequency. This method is below briefly presented:

The following abbreviations are proposed:

AV = **A**sset **V**alue – the costs needed for replacement and those caused by losses of any of the organization assets (tangible and intangible);

PVL = **P**otential **V**alue **L**oss – measures the impact on assets or quantifies the loss when the event occurs, being expressed in percentage (%) against the asset value;

ARO = **A**nalyzed **R**ate of **O**ccurrence – frequency the event is expected to occurred. It is a statistical estimation usually provided by some specialty institutions based on some data that it already has.

ALE is calculated according to the formula:

$$ALE = AV \times PVL \times ARO \qquad (3)$$

There can be included another measure – recalculated ALE – representing the maximum value of the security investments that we agree to spend for protection against threats.

*Calculation example:*

*ARO (frequency) – has value* $\frac{1}{10}$ *( the event can occur once every 10 years)*

*PVL (potential asset value loss) – 50%;*

AV (asset value) is considered at 50000 €;

ALE = AV x PVL x ARO = 50000 x 0.50 x 0.1 = 2500 €

In other words, there is the possibility that due to not taking into account the risk associated to this asset loss, the organization may loose every year 2500 €.

Sometimes another indicator is utilized – SOL (**S**ingle **O**ccurrence **L**oss) – when the company is interested on the value of the damage caused by a single occurrence of a threat in case of extremely important assets and the second occurrence does no longer count. It is similar to ALE, but the result of this analysis forms the response to the question "**It is possible to assure or not**" to continue the business in case of disaster.

*Calculation example*:

AV = 5 millions €;

PVL = 10%;

$SOL - AV \times PVL = 5 \times 10^6 \times 10^{-1} = 500000$ €.

Both methods may serve as basis for impact assessment, when all assets, weak points, threats and their occurrence frequencies have been identified. The impacts can be determined and consequently grouped into logical structured groups.

There are several such structure modalities, the basic example having only with 2 impact levels: "tolerable"(T) or "Non-Tolerable" (N).

The method recommended by ISO IEC TR 13335-3 is to complete a simple matrix, while considering the possible loss value and the threat occurrence frequency. For both variables there are decided 5 classes, the highest one being 4 as can be noticed in table 2.

**Table 2:** Impact assessment considering loss value and the threat occurrence frequency

| Possible losses | 0 | 1 | 2 | 3 | 4 |
|---|---|---|---|---|---|
| Frequency | | | | | |
| 4 | 0 | NT | NT | NT | NT |
| 3 | T | NT | NT | NT | NT |
| 2 | T | T | NT | NT | NT |
| 1 | T | T | T | NT | NT |
| 0 | T | T | T | T | 0 |

In this case the impact can be associated with the risk.

## 4. A case study about the evolution of the impact

Every year, Computer Security Institute (CSI) carries out different studies with the participation of the San Francisco Federal Bureau of Investigation's Computer Intrusion Squad.

These studies are used by every company to concentrate their efforts to counteract the most frequent attacks. The survey results are based on the opinion and the responses of different computer security practitioners in U.S. corporations, government agencies, financial institutions, medical institutions and universities.

The CSI survey it is said to have been conducted anonymously as a way of enabling respondents to speak freely about potentially serious and costly events that have occurred within their networks over the past year. Organizations covered by the survey include many areas from both the private and public sectors. The sectors with the largest number of responses came from the financial sector (20 percent), followed by consulting and education (11 percent), informatics technology (10 percent), and manufacturing (8 percent). The diversity of organizations responding was also reflected in the 9 percent designated as "Other" [6].

Taking into account the information given by CSI, we made comparisons between the last two years, 2006 and 2007 for distinguishing the losses suffered by the companies. The average annual loss reported in 2007 survey shot up to $350424 from $168000 the previous year. The result are presented in table 3.

**Table 3**: The evolution of the losses between 2006 and 2007

| Type of abuse | Value 2006 USD | Value 2007 USD | % |
|---|---|---|---|
| Financial Fraud | 2556900.00 | 21124750.00 | 7.26 |
| Virus | 15691460.00 | 8391800.00 | -0.47 |
| System penetration by outsider | 758000.00 | 6875000.00 | 8.07 |
| Theft of confidential data | not mentioned | 5685000.00 | |
| Laptop or mobile hardware theft | 6642660.00 | 3881150.00 | -0.42 |
| Insider abuse of Net access or email | 1849810.00 | 2889700.00 | 0.56 |
| Denial of services | 2922010.00 | 2888600.00 | -0.01 |
| Fishing | 647510.00 | 2752000.00 | 3.25 |
| Bots | 923700.00 | 2869600.00 | 2.11 |
| Theft of proprietary info from mobile device theft | 6034000.00 | 2345000.00 | -0.61 |
| Theft of confidential data from mobile device theft | not mentioned | 2203000.00 | |
| Sabotage of data or network | 260000.00 | 1056000.00 | 3.06 |
| Unauthorized access to informatics | 10617000.00 | 1042700.00 | -0.90 |
| Web site defacement | 162500.00 | 725000.00 | 3.46 |
| Telecom fraud | 1262410.00 | 651000.00 | -0.48 |
| Misuse of wireless network | 269500.00 | 542850.00 | 1.01 |
| Misuse of public application | not mentioned | 251000.00 | |
| Instant messaging abuse | 291510.00 | 200700.00 | -0.31 |
| Password sniffing | 161210.00 | 168100.00 | 0.04 |
| Blackmail | not mentioned | 160000.00 | |
| Exploit of yours organization DNS server | 90100.00 | 104500.00 | 0.16 |
| Abuse of wireless network | 469010.00 | not mentioned | |
| Others | 885000.00 | 123500.00 | -0.86 |
| **Total** | **52494290.00** | **66930950.00** | **0.28** |

At this survey participated in the year 2006, 616 computer security practitioners and in 2007 - 494. The results of this survey indicate that cyber crime is a critical concern. Every organization is vulnerable to numerous types of attack from many sources and the result of an intrusion can be devastating in terms of lost assets and good will.

Insider abuse of network access or email edged out virus incidents as the most prevalent security problem.

It is easily seen that the most important losses are because of the system penetration by the outsiders and the financial fraud. The losses caused by those factors are very import – the sums were raised with 8.07% and with 7.26%.

It is seen that the biggest loss in 2007 is due to financial fraud and not by viruses like in 2006. I cannot say that these results can be expanded to all the companies because of the small number of respondents but I can consider them important because the survey is sent to roughly the same group each year.

About the costs of computer crime – there are some informatics about those cost and the percentage of IT Budget Spent on Security. 61% said that their organizations allocated 5 % or less of their overall IT budget to informatics security like it is seen in the figure 3:

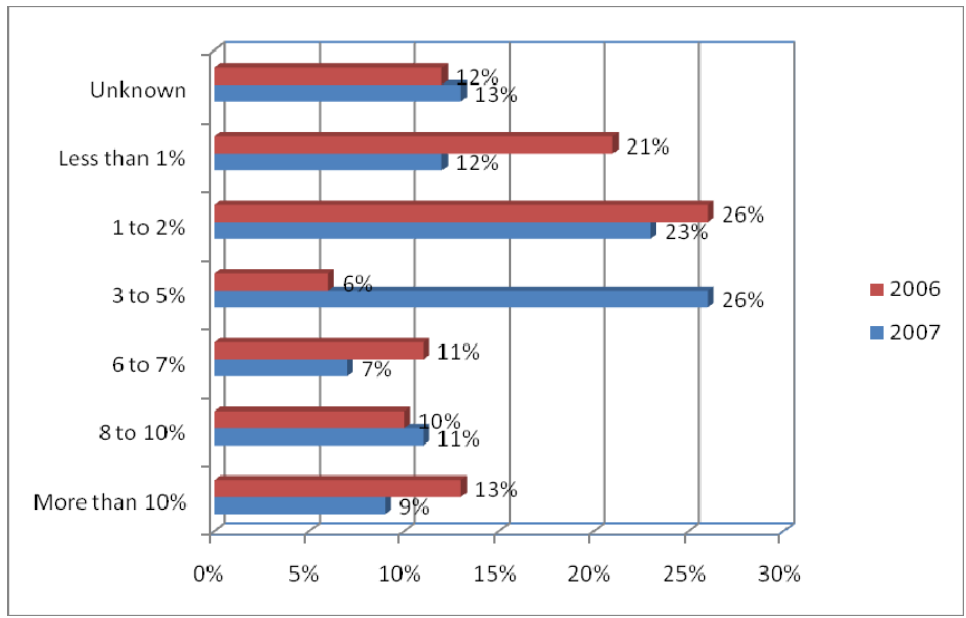

**Figure 3.** The percentage of IT Budget Spent on Security

A quick comparison of the bars at 3 to 5 % level shows a significant uptick in 2007 and we must notice that in 2006, 47 % said their organization allocated less than 3 % of the total IT Budget, whereas this year only 35 % fell into that range.

The general picture is that security programs budgets are slightly up. Expressing the budget as a percentage of the IT budgets means that the actual number of dollars spent depends on whether the IT budget is growing or decreasing.

**Conclusions**

Can say that security is improving within companies, but, the losses to cybercrime are bigger this year than the last one.

The country's economy relies on networked computer informatics systems for communications, energy distribution, commerce, transportation and in other domains. It is known that cybercrime and the attendant threat of identity theft reduce user and consumer confidence, reducing the acceptance of e-commerce.

And so, computer security has moved to a position of prominence in most organization being a critical activity that helps protecting the systems. But, we can say that if we want to have a certain security level we must first know the threat we are bewaring of. And after that, when testing the organization's security, we must act like a real hacker because in that way we can discover all the lacks in our system and then, periodically we must develop new vulnerability tests.